

\def\date{le\ {\the\day}\ \ifcase\month\or janvier\or
{f\'evrier}\or mars\or avril \or mai\or juin\or juillet\or
{ao\^ut}\or septembre\or octobre\or novembre\or {d\'ecembre}\fi
\ {\oldstyle\the\year}}

\let\noi=\noindent

\def\a{\alpha}

\def\ve{\varepsilon}

\def\lda{\lambda}

\def\D{\Delta}

\def\L{\Lambda}

\font\tenbb=msym10

\font\sevenbb=msym7
\font\fivebb=msym5

\newfam\bbfam
\textfont\bbfam=\tenbb \scriptfont\bbfam=\sevenbb
\scriptscriptfont\bbfam=\fivebb
\def\bb{\fam\bbfam}

\def\C{{\bb C}}

\def\un{{\rm 1\mkern-4mu  l }}

\font\titre=cmbx12

\font\got=eufm10

\def\part{\partial}

\def\ra{\rightarrow}

\def\sbs{\subset}

\def\and{\mathop{\rm and}\nolimits}

\def\id{\mathop{\rm id}\nolimits}

\def\Im{\mathop{\rm Im}\nolimits}

\catcode`\@=11
\def\Eqalign#1{\null\,\vcenter{\openup\jot\m@th\ialign{
\strut\hfil$\displaystyle{##}$&$\displaystyle{{}##}$\hfil
&&\quad\strut\hfil$\displaystyle{##}$&$\displaystyle{{}##}$
\hfil\crcr#1\crcr}}\,} \catcode`\@=12

\catcode`\@=11
\def\displaylinesno #1{\displ@y\halign{
\hbox to\displaywidth{$\@lign\hfil\displaystyle##\hfil$}&
\llap{$##$}\crcr#1\crcr}}

\def\ldisplaylinesno #1{\displ@y\halign{
\hbox to\displaywidth{$\@lign\hfil\displaystyle##\hfil$}&
\kern-\displaywidth\rlap{$##$}
\tabskip\displaywidth\crcr#1\crcr}}
\catcode`\@=12

\def\buildrel#1\over#2{\mathrel{
\mathop{\kern 0pt#2}\limits^{#1}}}

\def\build#1_#2^#3{\mathrel{
\mathop{\kern 0pt#1}\limits_{#2}^{#3}}}

\def\hfl#1#2{\smash{\mathop{\hbox to 6mm{\rightarrowfill}}
\limits^{\scriptstyle#1}_{\scriptstyle#2}}}

\def\vfl#1#2{\llap{$\scriptstyle #1$}\left\downarrow
\vbox to 3mm{}\right.\rlap{$\scriptstyle #2$}}

\def\vfll#1#2{\llap{$\scriptstyle #1$}\left\uparrow
\vbox to 3mm{}\right.\rlap{$\scriptstyle #2$}}

\def\diagram#1{\def\normalbaselines{\baselineskip=0pt
\lineskip=10pt\lineskiplimit=1pt}   \matrix{#1}}

\def\up#1{\raise 1ex\hbox{\sevenrm#1}}

\def\signed#1 (#2){{\unskip\nobreak\hfil\penalty 50
\hskip 2em\null\nobreak\hfil\sl#1\/ \rm(#2)
\parfillskip=0pt\finalhyphendemerits=0\par}}

\def\TeX{T\kern-.1667em\lower.5ex\hbox{E}\kern-.125em X}

\def\lsim{ {\raise -3mm \hbox{$<$} \atop \raise 2mm
\hbox{$\sim$}} }

\def\gsim{ {\raise -3mm \hbox{$>$} \atop \raise 2mm
\hbox{$\sim$}} }

\def\frac#1#2{\mathop{\scriptstyle#1\over\scriptstyle#2}\nolimits}

\def\fnote#1{\advance\noteno by 1\footnote{$^{\the\noteno}$}
{\eightpoint #1}}

\def\boxit#1#2{\setbox1=\hbox{\kern#1{#2}\kern#1}%
\dimen1=\ht1 \advance\dimen1 by #1 \dimen2=\dp1 \advance\dimen2 by
#1
\setbox1=\hbox{\vrule height\dimen1 depth\dimen2\box1\vrule}%
\setbox1=\vbox{\hrule\box1\hrule}%
\advance\dimen1 by .4pt \ht1=\dimen1
\advance\dimen2 by .4pt \dp1=\dimen2 \box1\relax}

\def\cube{
\raise 1 mm \hbox { $\boxit{3pt}{}$}
}

\def\cqfd{\unskip\kern 6pt\penalty 500
\raise -2pt\hbox{\vrule\vbox to10pt{\hrule width 4pt
\vfill\hrule}\vrule}\par}

\def\dstar {\displaystyle ({\raise- 2mm \hbox
{$*$} \atop \raise 2mm \hbox {$*$}})}

\def\ref #1#2{
\smallskip\parindent=1,0cm
\item{\hbox to\parindent{\enskip\lbrack{#1}\rbrack\hfill}}{#2} }

\def\choose#1#2{\mathop{\scriptstyle#1\choose\scriptstyle#2}\nolimits}

\def\adots{\mathinner{\mkern2mu\raise1pt\hbox{.}
\mkern3mu\raise4pt\hbox{.}\mkern1mu\raise7pt\hbox{.}}}

\def\pegal{\mathrel{\vbox{\hsize=9pt\hrule\kern1pt
\centerline {$\circ$}\kern.6pt\hrule}}}
\magnification=1200
\overfullrule=0mm

\def\Mat{\mathop{\rm Mat}\nolimits}

\centerline {\titre Tetramodules over the Hopf algebra of regular
functions on
a torus.}
\vglue 1cm
$$\displaylines{
{\rm Tanya\  Khovanova} \cr
\hbox{\rm Department of Mathematics } \cr
\hbox{\rm Massachusetts Institute of Technology} \cr
\hbox{\rm e-mail: tanyakh@math.mit.edu} \cr
\hbox{\rm Cambridge, MA 02139} \cr
}$$

\vglue 1.5cm
\noi {\bf Introduction.}
\medskip
The definition of a tetramodule appeared in my joint work
with Joseph Bernstein [B-KH]. In this paper we initiated
an axiomatic approach to the construction of the quantum group
$SL_q(2)$. We hope to use this approach more universally.
\medskip
The notion of a tetramodule seems to be interesting in itself.
The goal of this paper is to describe the basic properties of
tetramodules.
\medskip
I am thankful to Joseph Bernstein who encouraged me
to write this paper. I would like to thank Victor Kac and
David Kazhdan for helpful discussions.
\medskip
The author was supported by an NSF Grant \# DMS-9306018.
     \bigskip
\noi {\bf 1. Motivations.}
\medskip
{\bf 1.1.} Let  $S$  be  a Hopf algebra.
We would like to study pairs  $(A,I)$,
where $A$ is a Hopf algebra (with multiplication $m$ and
comultiplication $\D$) and $I  \subset  A$  a  two-sided  Hopf
ideal, such that $A/I$ is isomorphic to $S$ as a Hopf algebra.
\medskip
{\bf 1.2.} Note that the comultiplication $\D : A \ra A \otimes A$
leads
to two $S$-comodule structures on $A$:
$$\displaylines {c_\ell : A \ra S \otimes A \qquad c_\ell = (pr \otimes id) \D
\cr c_r : A \ra A \otimes S \qquad c_r = (id \otimes pr) \D \cr
}$$
where $pr$ is the natural projection $A \ra S = A/I$.
\medskip
{\bf 1.3.} Consider the associated graded algebra $gr\, A$:
$$gr\, A = \build{\oplus}_{0 \le n}^{} gr_n\ A, $$
where $$gr_n \,A = I^n/I^{n+1}\ .$$
It is easy to see that $gr\, A$ inherits the structure of a graded Hopf algebra
with $gr_0 A$ equal to $S$. In particular, $grA$ has the structure of
a graded $S$-bicomodule.
\medskip
{\bf 1.4.} We have two natural $S$-module structures on $grA$. These
structures commute and preserve $gr_n A$.
\medskip
The $S$-bicomodule and $S$-bimodule structures are compatible:
for any $s \in S,\ x \in gr A$
$$c_\ell(xs) = c_\ell(x) \D s\ .$$
The other three relations are of the same type:
$$c_\ell(sx) = \D s\cdot c_\ell(x)$$
$$c_r(xs) = c_r(x) \D s$$
$$c_r(sx) = \D s\cdot c_r(x).$$
\medskip
{\bf 1.5.} {\it Definition}. We call the linear space $V$ an $S$-{\it
tetramodule} if $V$ is equipped with commuting left and right $S$-module
structures, commuting left and right $S$-comodule structures, and
the $S$-bimodule and $S$-bicomodule structures are compatible (see 1.4).
\medskip
{\bf 1.6.} {\it Example}. Tetramodules first appeared in my work with
Joseph Bernstein [B-KH]; where
$A$ was the Hopf algebra of regular functions on the quantum group $SL_q(2)$,
and
$S$ was the Hopf algebra of regular functions on the
one-dimensional torus $H$.
In this case an $S$-comodule structure defines an algebraic
representation of $H$.
\bigskip
\noi {\bf 2. Definition of tetramodule.}
\medskip

{\bf 2.1.} Let us rewrite the definition of $S$-tetramodule.
\medskip
{\it Definition.} Given a Hopf algebra $S$, an $S$-tetramodule
is a vector space $V$ equipped with four morphisms
$$ \matrix {
m_\ell & : \quad  S \otimes V  & \ra \quad V            \hfill       \cr
m_r    & : \quad  V \otimes S  & \ra \quad V            \hfill       \cr
c_\ell & : \quad  V \hfill     & \ra \quad S  \otimes V              \cr
c_r    & : \quad  V \hfill     & \ra \quad V  \otimes S              \cr
}$$
satisfying the following relations -- $H1, H2, H3$:
\medskip
\noi $H1.$ The morphism $m_\ell$ (resp. $m_r$) defines
the structure of a left (resp.  right) $S$-module on $V$.  This means
that it is  associative,  and  the  element  $1 \in S$ acts as
the identity.
\medskip
$H1^\prime .$ The actions $m_\ell$ and $m_r$ commute on $V$.
This means  that the operator $m_r (m_\ell \otimes id)$ equals
the operator $m_\ell (id \otimes m_r)$ as an operator from $S
\otimes V \otimes S$
to $V$.
\medskip
\noi $H2.$ The morphism $c_\ell$ (resp. $c_r$) defines
the structure of left (resp. right) $S$-comodule on $V$. This means
that it is coassociative,  and the counit  acts  as
the identity.
\medskip
$H2^\prime .$ The  actions  $c_\ell$  and  $c_r$
cocommute on $V$. This   means   that   the   operator  $(id  \otimes
c_r)c_\ell$ from $V$ to $S \otimes V \otimes S$ equals  the
operator $(c_\ell \otimes id)c_r$.
\medskip
\noi $H3.$ The connection between the $S$-module  and the
$S$-comodule
structures:
\medskip
$H3rl.$ This axiom describes the compatibility of $m_r$ and
$c_\ell$:
The morphism $m_r : V \otimes S \ra V$ is a morphism of left $
S$-comodules. In other words the following diagram commutes:
$$ \diagram {
{}                                   & V       &   {}                      \cr
 m_r \nearrow {}                     & {}      &   \searrow c_\ell         \cr
V \otimes S                          & {}      &    S \otimes V            \cr
\vfl{}{c_\ell \otimes \D}            & {}      &    \vfll {}{m \otimes m_r}\cr
(S \otimes V) \otimes (S \otimes S)  & \hfl {S_{2,3}} {}  & (S \otimes S)
\otimes (V \otimes S) \cr
}$$
Note that  this  diagram is also equivalent to the requirement
that the morphism $c_\ell : V \ra S \otimes V$ is a morphism of right $
S$-modules.
\medskip
Similarly, we define the connection axioms $H3ll,  H3lr, H3rr$
describing the compatibility of
pairs   $(c_\ell,m_\ell),\   (c_r,m_\ell),\   (c_r,  m_r)$.
\bigskip
     \noi {\bf 3. Decomposition of tetramodule.}
\medskip
{\bf 3.1.}
{}From now on we consider only the case when  $S$  is  the  Hopf algebra  of
regular functions on a torus $H$.  In this case we can give an
explicit description  of the category of $S$-tetramodules [B-KH].
\medskip
{\bf 3.2.} We
use the following standard
\medskip
{\it Lemma.} Let $W$ be an $S$-module equipped with the compatible
algebraic action of the group $H$. Then $W = S \otimes W^H$, where $W^H$ is
the
space of $H$-invariants.
\medskip
{\bf 3.3.} Let us apply this lemma to our case. Let $V$ be an $S$-tetramodule.
Applying lemma 3.2 to the right action of $H$ on $V$ and the
right multiplication by $S$ we can write $V$ as
$V = V^H \otimes S$.
\medskip
{\bf 3.4.} Now let $V$ be an $S$-tetramodule $V = V^H \otimes S$. We want
to describe an $S$-tetramodule structure on $V$ in terms of some structures on
$V^H$.
\medskip
The right action of $H$ on $V^H$ is trivial. It is clear that $V^H$ is
$ad_H$-invariant, so the left action of $H$ on $V^H$ coincides with the $ad_H$
action. Hence, knowing the $ad_H$ action on $V^H$, we can reconstruct the left
and right actions of $H$ on $V$.
\medskip
The right action of $S$ on $V$ is defined by decomposition $V = V^H \otimes
S$. Now we have to reconstruct the left action of $S$ on $V$.
\medskip
Let $\L$ be the lattice of characters of $H$. Then $\L \subset S$ is a basis of
$S$. For $\lda \in \L$ consider operators $m_\ell(\lda)$ and
$m_r(\lda)$ of left and right multiplications by $\lda$ in $V$,
and set  $L(\lda)   =   m_\ell(\lda)   m_r(\lda)^{-1}$.   Then
operators $L(\lda)$
commute with  the  right and the left action of $H$ and hence
preserve the subspace $V^H$.
\medskip
So we have defined a homomorphism $L$ of $\L$ into automorphisms of $V^H$,
commuting with $ad_H$. Knowing  $L$ we can reconstruct the left action of $S$
on $V$.
\medskip
{\bf 3.5.}  {\it  Summary.}  Let  $S$  be  the Hopf algebra of
regular functions on a torus $H$. Then the functor $V \to V^H$
gives an equivalence of
the category of $S$-tetramodules with the category of
algebraic $H$-modules equipped with the commuting action $L$ of the lattice
$\L$ .
\bigskip
\noi {\bf 4. Tetramodules over $S \otimes S$.}
\medskip
{\bf 4.1.} Let $B$ be a Hopf algebra. We call an imbedding
$i: B \ra S$ a Hopf imbedding if it is closed under
multiplication: $i \cdot m_B = m_S \cdot (i \otimes i)$ and respects
comultiplication: $(i \otimes i) \cdot \D_B = \D_S \cdot i$.
\medskip
We call a projection $p: S \ra B$ a Hopf projection if it is closed under
comultiplication: $(p \otimes p) \cdot \D_S = \D_B \cdot p$ and
respects multiplication: $p \cdot m_S = m_B \cdot (p \otimes p)$.
\medskip
{\bf 4.2.} Let $V$ be an $S$-tetramodule. A Hopf imbedding
$i: B \ra S$ gives us a $B$-bimodule structure
on $V$. A Hopf projection $p: S \ra B$ gives us a $B$-bicomodule
structure on $V$.
\medskip
{\it Theorem.} If $p \cdot i = {\rm id}$, then
these bimodule and bicomodule structures are compatible.
\medskip
{\it Proof.} Let us prove, for instance, the compatibility of
right $B$-module and right $B$-comodule structures on $V$.
The right $B$-comodule structure on $V$ is defined as
the composite map:
$$ V \hfl {c_r} {} V \otimes S \hfl {id \otimes p} {} V \otimes B.$$
We have to prove that this composition defines a homomorphism of
$B$-modules. The first map defines a homomorphism of $S$-modules,
and, hence, of $B$-modules.
\medskip
The $B$-module structure on $V \otimes B$ is defined as
$(i \otimes id) \cdot \D_B$. The $B$-module structure on
$V \otimes S$ is defined as $\D_S \cdot i$, which is equal
to $(i \otimes i) \cdot \D_B$ for Hopf imbedding $i$.
So, all is left to prove is that
$p: S \ra B$ defines a homomorphism of $B$-modules.
It is by definition of $p$ that $p$ defines a
homomorphism of $S$-modules. The induced $B$-module
structure on $B$ is equal to $p \cdot i(B)$, which is equal
to the existing $B$-module structure on $B$.
\medskip
{\bf 4.3.} Denote by $\eta$ the unit in $S$: $\ \eta: k \ra S$, and by
$\ve$ the counit in $S$: $\ \ve: S \ra k$.
There are three natural Hopf imbeddings $S \ra S \otimes S$:
\medskip
(i) left - $i_{\ell}: S = S \otimes k \hfl {id \otimes \eta} {}
S \otimes S$;
\medskip
(ii) right - $i_r: S = k \otimes S \hfl {\eta \otimes id} {}
S \otimes S$;
\medskip
(iii) comultiplication - $\D$.
\medskip
There are three natural Hopf projections $S \otimes S \ra S$:
\medskip
(i) left - $p_{\ell}: S \otimes S \hfl {id \otimes \ve} {}
S \otimes k = S$;
\medskip
(ii) right - $p_r: S \otimes S \hfl {\ve \otimes id} {}
k \otimes S = S$;
\medskip
(iii) multiplication - $m$.
\medskip
{\bf 4.4.} Using the theorem it is easy to check which pairs
of structures are compatible. The results are shown in the
following table, where plus marks the compatibility:
$$ \matrix {
{} & i_{\ell} & i_r & \D \cr
p_{\ell} & + & - & + \cr
p_r      & - & + & + \cr
m        & + & + & - \cr
}$$
\medskip
{\bf 4.5.} Let $V_1, V_2$ be two $S$-tetramodules. The space
$V_1 \otimes V_2$ has the natural structure of an
($S \otimes S$)-tetramodule.
Using the discussion above, we can introduce various $S$-tetramodule
structures on $V_1 \otimes V_2$. Our goal is to give a
natural definition of tensor product in the category
of tetramodules. From this point of view the space $V_1 \otimes
V_2$ is "too big". Its $(S \otimes S)$-tetramodule structure
is "$S$ times too much" for an $S$-tetramodule.
The definition of the tensor product is given in the next section.
\bigskip
\noi {\bf 5. Tensor products of $\bf S$-tetramodules.}
\medskip
{\bf 5.1.} Let $V_1, \ V_2$ be two $S$-tetramodules. Denote $W = V_1
\otimes_S V_2$. We introduce an $S$-bimodule structure on $W$ by
following formulas:

$$\Eqalign {
m_\ell &: &S \otimes W        & \ra W                      \cr
{}     &  &(f,v_1 \otimes v_2) &\mapsto (f v_1 \otimes v_2) \cr
\cr
m_r    & : &W \otimes S          &\ra W                       \cr
{}     &   &(v_1 \otimes v_2,f)  &\mapsto (v_1 \otimes v_2 f) \cr
}$$
and an $S$-bicomodule structure by:
$$
\eqalign {
s_\ell(h) (v_1 \otimes v_2) &= s_\ell(h) v_1 \otimes s_\ell
(h) v_2 \cr
s_r (h) (v_1 \otimes v_2) &= s_r (h) v_1 \otimes s_r (h) v_2 \ ,\cr
}$$
where $s_l(h) \ (s_r(h))$ is the left (right) action of the point
$h$ of the torus $H$.
\medskip {\it Statement.} These $S$-bimodule and $S$-bicomodule structures are
correctly defined and compatible.
\medskip
Therefore, $W = V_1 \otimes_S V_2$ is equipped with the natural
$S$-tetramodule structure.
\medskip
{\bf 5.2.} Let $V_1^H$ and $V_2^H$ be the spaces of right $H$-invariants in
$V_1$ and $V_2$. Then $W = V_1 \otimes_SV_2$ is
isomorphic to $V_1^H \otimes V_2^H \otimes S$. This isomorphism could be
realized through the map:
$$V_1^H \otimes V_2^H \otimes S \ra (V_1^H \otimes \un) \otimes (V_2^H
\otimes S) \sbs V_1 \otimes V_2 \ra V_1 \otimes_S V_2 \ .$$
The $S$-tetramodule structure on $W$ can be described as follows: $W^H
\approx V_1^H \otimes V_2^H$ is the space of right $H$-invariants in $W$.
The adjoint action of $H$ on $W^H$ equals
$$ad_H \mid _{V_1} \otimes \, ad_H \mid_{V_2}$$
and the operator $L(\lda)$ on $W^H$ equals
$$L(\lda) \mid_{V_1} \otimes\,  L(\lda) \mid_{V_2} \ .$$
\medskip
{\bf 5.3.} Thus the category of $S$-tetramodules is a monoidal
category and is equivalent to the monoidal category of linear spaces equipped
with an algebraic action of $H$ and a commuting action of $\L$.
\medskip
{\bf 5.4.} Let us introduce another tensor product.
Let $V_1,V_2$ be two $S$-tetramodules. Denote by $V
_1 \otimes ^S V_2$ a subspace
in $V_1  \otimes V_2$ of vectors $(v_1,v_2)$ such that, for any
$h \in H$:
$$s_r(h)v_1 \otimes s_\ell^{-1} (h) v_2 = v_1 \otimes v_2\ .$$
We introduce an $S$-bimodule structure on $W = V_1 \otimes ^S V_2$
by the following
formulas:
$$\Eqalign {
m_\ell &: &S \otimes W        & \ra W                      \cr
{}     &  &(f, v_1 \otimes v_2) &\mapsto \D f \cdot( v_1 \otimes v_2) \cr
\cr
m_r    & : &W \otimes S          &\ra W                       \cr
{}     &   &(v_1 \otimes v_2,f)  &\mapsto (v_1 \otimes v_2) \D f \cr
}$$
and an $S$-bicomodule structure by:
$$\eqalign {
s_\ell (h) (v_1 \otimes v_2) &= s_\ell(h) v_1 \otimes v_2 \cr
s_r (h) (v_1 \otimes v_2) &= v_1 \otimes s_r (h) v_2 \ . \cr
}$$
\medskip {\it Statement}. These $S$-bimodule and $S$-bicomodule structures are
correctly defined and compatible.
\medskip
So $W$ is equipped with the natural $S$-tetramodule structure.
\medskip
{\bf 5.5.} {\it Lemma.} $V_1 \otimes_S V_2 $ and $V_1
\otimes^S V_2$ are canonically isomorphic.
\medskip
{\it Proof.} Let $V_1^H$ be the space of right $H$-invariants in $V_1$ and
${}^HV_2$ be the space of left $H$-invariants in $V_2$. Then $V_1 = V_1^H
\otimes S$ and $V_2 = S \otimes {}^H V_2$. A natural projection $V_1
\otimes V_2 \ra V_1 \otimes_S V_2 = V_1^H \otimes S \otimes {}
^H V_2$ is given by
$$V_1^H \otimes (S \otimes S) \otimes {}^H V_2
\build{\longrightarrow}_{}^{\id \otimes m \otimes \id} V_1^H \otimes S
\otimes {}^H V_2.$$
     We can describe the induced $S$-tetramodule structure on $V_1
^H \otimes S \otimes {}^H V_2$ as follows.  $V_1^H  \otimes  S
\otimes {}^H   V_2   =  V_1  \otimes  {}^H  V_2$,  hence the left
$S$-module and $S$-comodule structures on $V_1$  define  left
structures on    $V_1^H   \otimes   S   \otimes   {}^H   V_2$.
Symmetrically, $V_1^H \otimes  S  \otimes  {}^H  V_2  =  V_1^H
\otimes V_2$, hence the right structures on $V_2$ define right
structures on $V_1^H \otimes S \otimes {}^H V_2$.

     We have the natural imbedding $V_1^H \otimes S \otimes {}
^H V_2 \ra V_1 \otimes
V_2$
$$V_1^H \otimes S \otimes {}^HV_2 \build {\longrightarrow}_{}^{\id \otimes
\D \otimes \id} (V_1^H \otimes S) \otimes (S \otimes \, {}^HV_
2)\ .$$
     It is easy to check that  this  imbedding  gives  us an
isomorphism of $S$-teramodules $V_1^H \otimes S \otimes {}^H V_2$
and $V_1 \otimes^S V_2$.
So $V_1 \otimes_S V_2$ and $V_1 \otimes^S V_2$
are both canonically isomorphic to $V_1^H \otimes S \otimes {}^H V_2$.
\medskip
{\bf 5.6.} {\it Remark.} $V_1 \otimes_S V_2$ and $V_1
\otimes^S V_2$ are canonically isomorphic, but their
definitions seem to be different. These definitions are dual in some sense
which we will not discuss here.
\bigskip
\noi {\bf 6. Universal graded Hopf algebra.}
\medskip
{\bf 6.1.} Let us return to a Hopf algebra $A$ with a Hopf ideal $I$,
such that $gr_0 A$ equals $S$. We denote $gr_1 A$ by $T$.
Algebra $gr A$ is generated by $S \oplus T$.
\medskip
{\bf 6.2.} {\it Lemma.} Given an $S$-tetramodule $T$,  there exist a graded
Hopf algebra
$\tilde A(S,T)$,  such  that
$\tilde A_0
= S,\  \tilde A_1 = T$, $\tilde A$ supplies $T$ with the given
$S$-tetramodule structure;  and  $\tilde  A$ is universal with
respect to these properties. The Hopf algebra $\tilde A$ is defined
up to a canonical isomorphism.
\medskip
{\bf 6.3.}  Explicitly,  the  universal  Hopf  algebra  $\tilde
A(S,T) = \oplus  \tilde  A_n$  can  be described as follows:
$\tilde A_n$ equals $T \otimes_S T \otimes_S \dots \otimes_S  T$ ($n$
factors) and
has an $S$-tetramodule structure described in 5. The
multiplication is natural:
$$(\tilde A_n, \tilde A_m) \ra \tilde A_n \otimes_S \tilde
A_m = \tilde A_{n+m}\ .$$
\medskip
{\bf 6.4.} For describing the comultiplication we use the fact
that $\tilde A \otimes \tilde
A$ is the graded algebra
$$ (\tilde A \otimes \tilde A)_n = \build{\oplus}_{i = 0}^{n} (
\tilde A_i \otimes \tilde A_{n-i})\ ;$$
and we already have the comultiplication formula for $\tilde A_0$ and $
\tilde A_1$:
$$\Eqalign{
\D : \tilde A_0 & \ra (\tilde A \otimes \tilde A)_0 \cr
\D : S &\ra S \otimes S \cr
\D s &= s \otimes s \cr
\cr
\D : \tilde A_1 & \ra (\tilde A \otimes \tilde A)_1 \cr
\D : T & \ra S \otimes T + T \otimes S \cr
\D t &= c_\ell(t) + c_r (t) \ . \cr
}$$
The Hopf algebra $\tilde A$ is generated by $\tilde A_0 \oplus
\tilde A_1$, so using the fact that the comultiplication is
a morphism   of   algebras,   we   can  easily  calculate  the
comultiplication formula for any element of $A$:
$\D(t_1 \otimes t_2)$ is equal to $\D t_1 \cdot \D t_2$ and so on.
\medskip
It is easy to prove that this definition is correct and supplies the algebra
$\tilde A$ with the bialgebra structure.
\medskip
{\bf 6.5.} {\it Antipode.} There is an antipode on $S$.
Using the following  commuting diagram
$$ \diagram {
S \otimes T + T \otimes S & \hfl {i \otimes id} {} &
S \otimes T + T \otimes S \cr
\vfll {} {\D} & {} & \vfl {} {m} \cr
T & \hfl {\eta \ve = 0} {} &  T \cr
} $$
we can easily define an antipode on $T$ and by induction an antipode
on $\tilde A$. Thus the bialgebra $\tilde A$ is supplied with
the Hopf algebra structure.
\medskip
{\bf 6.6.} {\it Remark.} The space $\tilde A \otimes^S \tilde A$
is a subspace in $\tilde A \otimes \tilde A$. It is easy to check
that $\Im \D \in \tilde A \otimes^S
 \tilde A$.
\medskip
{\bf 6.7.}
If $A$ is a Hopf algebra which corresponds to the same $S$-tetramodule $T$,
then we have a natural morphism of Hopf algebras:
$$\tilde A (S,T) \ra gr\, A\ .$$
\bigskip
\noi {\bf 7. Examples.}
\medskip
{\bf 7.1.} Below we list some examples of a Hopf algebra $A$
with a natural Hopf ideal $I$, $\ A/I = S$. We describe an $S$-tetramodule
$T = I/I^2$.
\medskip
{\bf 7.2.} {\it Lie case.} Let $G$ be a reductive algebraic group,
$H$ its Cartan subgroup. Let $A = \C[G]$ be the Hopf algebra of regular
functions on $G$ and $I$ an ideal of functions equal to $0$ on $H$.
Then $S = A/I$ equals $\C[H]$, $T = I/I^2$ is an $S$-tetramodule.
The space $T^H$ is isomorphic to $(\got g/\got h)^*$.
\medskip
As an $H$-module, $T^H$ is a direct sum of one-dimensional representations
$T_\a^H$ which correspond to roots of $G$. An $S$-bimodule structure of $V$
is trivial, which is equivalent to the fact that the lattice $\lambda$
acts on $T^H$ as the identity.
\medskip
{\bf 7.3.} {\it $SL_q(n)$.} The algebra $A$ of functions on $SL_q(n)$ is
defined as an algebra generated by $n^2$ noncommuting elements
$a_{ij} \ (1 \le i,j \le n)$, satisfying the following relations [M]:
\medskip
Introduce matrices
$$ \Eqalign {
Y(i,j,k,l) & = \pmatrix {a_{ij} & a_{il} \cr a_{kj} & a_{kl} \cr}
\cr \cr
Q & = \pmatrix {0 & -1 \cr q^{-1} & 0 \cr}.
}$$
\medskip
Then for every $1 \le i < k \le n$, $1 \le j < l \le n$
we put relations:
$$\matrix {
\eqalign {Y Q Y^t &= x_1 Q \cr
Y^t Q Y &= x_2 Q \cr} &\quad x_1,x_2 \in \C^* \cr
}.$$
\medskip
We have one more relation for the determinant:
$$\build {\Sigma}_{s \in S_n}^{} (-q)^{-l(s)} a_{1 \ s(1)}
\cdots a_{n \ s(n)} = 1.$$
\medskip
To define the comultiplication in $A$ we consider the $n \times n$
matrix $Z = \{ a_{ij} \}$. Using the natural imbeddings
$i', \ i'' : A \ra A \otimes A, \ \bigl ( i' (x) = x \otimes 1,\
i''(x) = 1 \otimes x \bigr )$, we can write the comultiplicaton
formulas as:
$$\D(Z) = i'(Z) \cdot i''(Z)\ ,$$
which is an equality in $\Mat (n, A \otimes A)$.
\medskip
The ideal $I$ is generated by $a_{ij} \ \ (i \ne j)$.
Then $A/I$ is isomorphic to $\C [H]$, where $H$ is an $(n-1)$-dimensional
torus. We define $T$ equal to $I/I^2$. It is easy to see that
$T^H$ is a direct sum of one-dimensional representations $T^H_\a$,
where $\a$ or $-\a$ is a simple root.
\medskip
It is easy to check that a character $\lda \in \L$ acts on $T_\a$
multiplyingit by $q^{-\a (\lda)} \ $ for $\a > 0$ and $q^{\a (\lda)} \ $
for $\a < 0$.
\medskip
{\bf 7.4.} {\it Quantum groups of $SL(2)$-type.} In the
paper [B-KH1] we constructed quantum groups of $SL(2)$-type.
Namely, these groups where attached to $S$ --  the space of regular
functions on the one-dimensional torus and an $S$-tetramodule $T$.
The space $T^H$ of $H$-invariants is two-dimensional:
$T^H = T^H_\a \oplus T^H_{-\a}$ (weight of $\a$ is equal to $n$)
and a character $s$ (the basis in $S$) acts on $T^H_\a$
multiplying it by $q_\a$, where $q^n_\a = q^n_{-\a}$.
\medskip
The corresponding Hopf algebra is generated by elements
$\hat h, \ h \in H \approx \C ^*$ and by elements $E$ and $F$
satisfying the relations:
$$\Eqalign{
\hat h_1 \hat h_2 & = \widehat {h_1h_2} \cr
\hat h E & = \a (h)E\hat h = h^nE\hat h \cr
\hat h F & = - \a (h) F \hat h = h^{-n} F \hat h \cr
[E,F] & = \frac {\hat q_\a - \hat q_{-\a}^{-1}}
{q_\a - q_{-\a}^{-1}} \cr
\D \hat h & = \hat h \otimes \hat h \cr
\D E & = \hat q_{\a} \otimes E + E \otimes 1 \cr
\D F & = 1 \otimes F + F \otimes \hat q_{-\a}^{-1}.
}$$
\bigskip
\noi {\bf References}
\medskip
[B-KH] J.Bernstein, T.Khovanova, \it On quantum group $SL_q(2)$,
\rm to be published.
\medskip
[L] G.Lusztig,  \it On quantum groups,  \rm J. of Alg.  \bf 131
\rm (1990).
\medskip
[M] Yu.I.Manin,  \it  Quantum   groups   and   non-commutative
geometry, \rm CRM, Universit\'e de Montr\'eal, 1988.
                           \end